# Diesel Generator Model Parameterization for Microgrid Simulation Using Hybrid Box-Constrained Levenberg-Marquardt Algorithm

Qian Long, *Student Member, IEEE*, Hui Yu, *Student Member, IEEE*, Fuhong Xie, *Student Member, IEEE,* Ning Lu, *Senior Member, IEEE* and David Lubkeman, *Fellow, IEEE*

*Abstract*—Existing generator parameterization methods, typically developed for large turbine generator units, are difficult to apply to small kW-level diesel generators in microgrid applications. This paper presents a model parameterization method that estimates a complete set of kW-level diesel generator parameters simultaneously using only load-step-change tests with limited measurement points. This method provides a more cost-efficient and robust approach to achieve high-fidelity modeling of diesel generators for microgrid dynamic simulation. A two-stage hybrid box-constrained Levenberg-Marquardt (H-BCLM) algorithm is developed to search the optimal parameter set given the parameter bounds. A heuristic algorithm, namely Generalized Opposition-based Learning Genetic Algorithm (GOL-GA), is applied to identify proper initial estimates at the first stage, followed by a modified Levenberg-Marquardt algorithm designed to fine tune the solution based on the first-stage result. The proposed method is validated against dynamic simulation of a diesel generator model and field measurements from a 16kW diesel generator unit.

## I. Introduction

Diesel generators are commonly used in hybrid microgrids, in conjunction with inverter-interfaced photovoltaic (PV), wind, and energy storage systems [1-3]. Therefore, it is critical to understand the dynamic response associated with control interactions between conventional diesel generators and inverter-based distributed energy resources (DERs). However, in many studies [4, 5] on control interactions between diesel generators and other DERs, modeling parameters of the diesel generator are arbitrarily obtained from publicly available data sources so it is not guaranteed that the dynamic response of the diesel generator will match field operation results under given operation conditions. In recent years, the hardware-in-the-loop (HIL) simulations are increasingly used to replace field tests of microgrid protection and control schemes [6, 7]. HIL models which are not sufficiently validated can lead to low-fidelity testing results [8]. Therefore, model parameterization method is needed for developing high-fidelity microgrid dynamic simulation.

Generator power plant model validation has been well established for power system dynamic studies [9-14]. However, the existing generator parameterization methods are mainly applicable to larger-scale steam-turbine generators and several technical challenges arise when these approaches are applied to the small-scale diesel generator model in microgrid applications. *First*, the existing methods are based on standard testing data or online recording measurements, collected from a series of sophisticated testing procedures, such as short-circuit tests, open-circuit tests, load rejection tests, frequency response tests, and so on [10-14]. This model validation process requires too large investment for small commercial microgrid owners. Thus, it is important to limit the scale, duration, and complexity of the field tests. *Second*, the existing methods heavily rely on the information from manufacturer datasheet such as equipment parameters, factory test results, and electromagnetic transient (EMT) benchmark models [9, 13]. For example, the generator parameters from manufacturer datasheet are needed as initial estimates for the model parameters. However, the manufacturer datasheet of the small-scale diesel generators does not contain as much information as that of multi-MW generator sets. Therefore, it is also desirable that the developed method is free from initial estimate selection. *Third*, the parameterization process needs to be automated with clearly defined performance metrics, such as accuracy, applicable operation conditions and parameter ranges. Thus, the manual approach from only trial-and-error or expert knowledge [13] is not applicable to the diesel generator model parameterization.

More advanced techniques have been recently proposed for model parameterization [15-19]. A model parameterization approach using Gauss-Newton algorithm is applied to solve the parameter estimation problem for power plant models [15]. However, the Gauss-Newton algorithm can encounter numerical difficulties for rank-deficient problems. The trust-region-reflective (TRR) based algorithms [16, 17] are chosen to solve parameter estimation problems with improved numerical robustness, but the performance of these algorithms depends on initial solution selection. There are researchers applying pure heuristic algorithms to parameter estimation problems. A new model identification method is introduced in [18] using a hybrid Cuckoo search algorithm to parameterize standard gas turbine and exciter models. A Particle-Swarm-Optimization based strategy is applied for key parameter estimation of permanent magnet synchronous machines with the inverter interface considered [19]. Although the authors mention that their methods can work with nonlinear models, it is noted that these heuristic methods are easily subject to long convergence time and premature convergence when dealing with nonlinear models.

To overcome the limited capability of the above-mentioned techniques, a new parameterization method using a hybrid box-constrained Levenberg-Marquardt (H-BCLM) algorithm is proposed to automatically estimate a complete parameter set for diesel generator models. Although a few hybrid algorithms are already proposed for model parameterization [16, 20], these algorithms are tailored to fit into PV and battery models that only contain algebraic equations. In this paper, the diesel generator model parameterization problem is formulated as a nonlinear least squares (NLSQ) problem for dynamic system. Only field measurements from load-step-change tests are used as the inputs of the algorithm. The proposed H-BCLM algorithm, which consists of two stages of solution searching process, combines the advantages of the heuristic approach and the optimization-based approach. A heuristic algorithm is applied at the first stage to find reasonable initial solutions within the predefined bounds, making the algorithm independent of manufacturer datasheet. These initial solutions are then used by an optimization-based algorithm to initiate the second-stage solution searching process. Because this NLSQ problem is rank-deficient, the Levenberg-Marquardt (LM) algorithm is selected in this paper to improve the robustness of the second-stage solution searching. Specifically, a variant of LM algorithm, called the box-constrained Levenberg-Marquardt (BCLM) algorithm, is implemented as the second-stage solution searching algorithm to solve the NLSQ problem in consideration of physical bounds of the parameters.

The rest of this paper is organized as follows. Section II describes the modeling method for diesel generator and the modeling parameters. Section III discusses the problem formulation and NLSQ for dynamic system. Section IV introduces the proposed H-BCLM algorithm. Section V presents the validation results, and conclusions and future work are provided in Section VI.

## II. Modeling Methodologies

### A. Diesel Generator Model

As shown in Figs. 1-3, the diesel generator model consists of multiple subsystems: diesel engine, excitation system, and synchronous generator. The model parameters are listed in Table I.

In this paper, the simplified diesel engine model introduced in [5] is used. As shown in Fig. 1, the model contains two elements: a proportional speed control and a mechanical actuator system. If zero steady-state error is required, the proportional speed control can be replaced by a proportional-integral-derivative controller. The dynamic model of the diesel engine can be expressed as:

$$\begin{cases} \dot{q}_1 = q_2 \\ \dot{q}_2 = -\frac{1}{T_2 T_3} q_1 - \frac{T_2 + T_3}{T_2 T_3} q_2 + P_{ref} + m(\omega_{ref} - \omega) \end{cases} \quad (1)$$

$$P_m = \frac{1}{T_2 T_3} q_1 + \frac{T_1}{T_2 T_3} q_2 \quad (2)$$

where $q_1$ and $q_2$ are state variables in the controllable canonical form of the diesel engine; $P_m$ is the mechanical power; $\omega$ is the rotor speed.

As shown in Fig. 2, the excitation system model [21], consists of an automatic voltage regulator (AVR) and an exciter. The AVR is modeled as a first-order transfer function while the exciter is modeled as a proportional-integral controller. A saturation block is added after the exciter output to avoid the excitation voltage from reaching an infeasible range. The excitation system model is expressed as:

$$\begin{cases} \dot{\xi}_1 = \xi_2 \\ \dot{\xi}_2 = -\frac{1}{T_V} \xi_2 + (V_{tref} - V_t) \end{cases} \quad (3)$$

$$V_f = \frac{K_V K_{ie}}{T_V} \xi_1 + \frac{K_V K_{pe}}{T_V} \xi_2 \quad (4)$$

where $\xi_1$ and $\xi_2$ are state variables in the controllable canonical form of the excitation system; $V_t$ are measured stator voltage magnitude; $V_f$ is the excitation voltage.

TABLE I
DIESEL GENERATOR MODEL PARAMETERS

| Symbol | Description | Units |
|---|---|---|
| $m$ | Speed droop gain | unitless |
| $\omega_{ref}$ | Speed reference | p.u. |
| $P_{ref}$ | Power reference | p.u. |
| $T_1, T_2, T_3$ | Diesel engine time constants | sec |
| $V_{tref}$ | Terminal voltage reference | p.u. |
| $T_V$ | AVR gain | sec |
| $K_V$ | AVR time constant | unitless |
| $K_{pe}, K_{ie}$ | PI gains for the exciter | unitless |
| $H$ | Inertia coefficient | sec |
| $\omega_s$ | Synchronous speed | p.u. |
| $D_f$ | Friction factor | p.u. |
| $X_d$ | d-axis synchronous reactance | p.u. |
| $X'_d$ | d-axis transient reactance | p.u. |
| $X_q$ | q-axis synchronous reactance | p.u. |
| $T'_{do}$ | d-axis transient open-circuit time constant | sec |
| $R_s$ | Stator resistance | p.u. |

The synchronous generator model is built based on the flux decay model in [22], as shown in Fig. 3. It is derived by eliminating fast damper winding dynamics $T'_{qo}$ when the inequality condition $T'_{qo} \ll T'_{do}$ holds true for the generator. The synchronous generator model includes both field winding dynamics and machine dynamics, and the equations are written as a third-order system as follows:

$$\begin{cases} \frac{2H}{\omega_s} \dot{\omega} = P_m - E'_q I_q - (X_q - X'_d) I_d I_q - D_f \omega \\ T'_{d0} \dot{E}'_q = -E'_q - (X_d - X'_d) I_d + V_f \\ \dot{\delta} = \omega - \omega_s \end{cases} \quad (5)$$

where $E'_q$ is the field flux linkage; $I_d$ and $I_q$ are d-axis and q-axis stator currents; $\delta$ is the rotor angular position.

The diesel generator model (1-5) is a seventh-order nonlinear dynamic system. Note that there are also network constraints that are written as algebraic equations. However, all the algebraic equations can be eliminated by substituting them into the differential equations. The form of the algebraic equations depends on the characteristics of the network. In a typical test setup, the network constraints simply represent one component, such as a source, a passive load or an induction motor. Take resistive loads for example, where the algebraic equations are written as follows:

$$\begin{bmatrix} 0 \\ E'_q \end{bmatrix} = \begin{bmatrix} R_s + R_{load} & -X_q \\ X'_d & R_s + R_{load} \end{bmatrix} \begin{bmatrix} I_d \\ I_q \end{bmatrix} \quad (6)$$

$$\begin{bmatrix} V_d \\ V_q \end{bmatrix} = \begin{bmatrix} R_{load} & 0 \\ 0 & R_{load} \end{bmatrix} \begin{bmatrix} I_d \\ I_q \end{bmatrix} \quad (7)$$

$$V_t = \sqrt{V_d^2 + V_q^2} \quad (8)$$

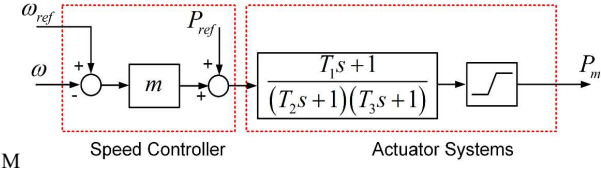

Fig. 1. Diesel engine model diagram

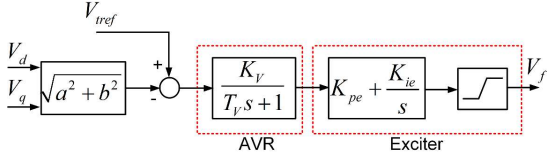

Fig. 2. Excitation system model diagram

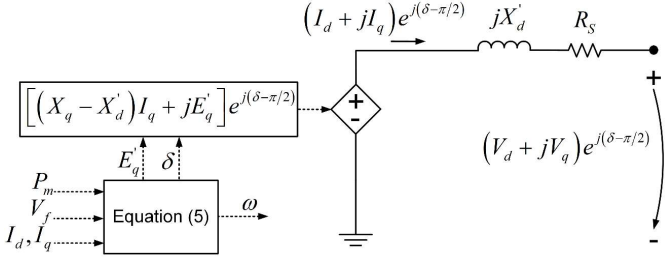

Fig. 3. Synchronous generator model one-line diagram

### B. Model Parameters

Table I lists all the parameters of the diesel generator model proposed in Section II-A. To find proper values for this complete set of parameters is difficult, especially for small-scale diesel generators. It is because the manufacturer datasheet of small-scale diesel generators might not be as informative and reliable as of the MW-level generators. Besides, even though the parameters are made available in the manufacturer datasheet, the accuracy of the parameters is another concern. For example, diesel generator inertia constant is such a parameter that is mostly either unavailable or inaccurate in the manufacturer datasheet. Therefore, the proposed model parameterization method is useful in this situation to get a decent parameter set for diesel generator modeling given only load-step-change test measurements.

A practical aspect of the model parameterization process is to decide parameter bounds. The choices of parameter bounds rely on prior information and knowledge. When there is no such expert knowledge, lower and upper bounds can be chosen intuitively with relaxed ranges. There are cases where manufacturer datasheets or expert knowledge are available for certain parameters. For example, the data are likely to be available for reactance of the generator, such as *dq*-axis synchronous and transient reactance $X_d$, $X_q$ and $X'_d$. Then these parameters are allowed to change within tight bounds while the rest of the parameters are free to vary in a significant range. How to incorporate parameter bounds into parameterization process is discussed in detail in Section IV.

## III. DIESEL GENERATOR MODEL PARAMETERIZATION PROBLEM FORMULATION

### A. Problem Formulation

The goal of the model parameterization problem is to find an optimal set of model parameters such that the mismatch between model response and the given measurements is minimized. The problem falls into the category of NLSQ problems, whose objective is defined as follows:

$$\min_{\boldsymbol{\theta}} h(\boldsymbol{\theta}) = \min_{\boldsymbol{\theta}} \frac{1}{2} \sum_{k=1}^{N} \|\boldsymbol{z}(t_k) - \boldsymbol{y}(t_k)\|_2^2 \quad (9)$$

where $\boldsymbol{\theta}$ is the vector in $\mathbb{R}^n$ containing model parameters to be estimated and $n$ is the length of the system parameter vector; $\boldsymbol{z}(t_k)$ and $\boldsymbol{y}(t_k)$ are the measured and simulated output vectors at time $t_k$, respectively, and both are in $\mathbb{R}^m$; $m$ is the length of the system output vector; $N$ is the number of data points.

The objective function (9) is subject to the nonlinear dynamic system equations defined above for the diesel generator model (1-8), with general representation written as follows:

$$\dot{\boldsymbol{x}}(t_k) = \boldsymbol{f}[\boldsymbol{x}(t_k), \boldsymbol{u}(t_k), \boldsymbol{\theta}], \boldsymbol{x}(t_0) = \boldsymbol{x_0} \quad (10)$$
$$\boldsymbol{y}(t_k) = \boldsymbol{g}[\boldsymbol{x}(t_k), \boldsymbol{u}(t_k), \boldsymbol{\theta}] \quad (11)$$

where $\boldsymbol{x}(t_k)$ is the state variable vector; $\boldsymbol{u}(t_k)$ is the input disturbance vector; $\boldsymbol{x_0}$ is the state variable vector with initial conditions; $\boldsymbol{y}(t_k)$ is the output vector. The functions $\boldsymbol{f}$ and $\boldsymbol{g}$ are nonlinear and real-valued. To be consistent with network constraints (6-8), we let $\boldsymbol{u}(t_k) = R_{load}(t_k)$ to represent the resistive load connected to the diesel generator, and it is a step signal when load step test is applied.

The diesel generator parameter bounds are defined in advance as follows:

$$\underline{\theta_i} \leq \theta_i \leq \overline{\theta_i}; i = 1, \dots, n \quad (12)$$

where $\underline{\theta_i}$ and $\overline{\theta_i}$ are the lower and upper bounds of $\theta_i$. The parameter bounds should be set to be consistent with the physical meanings of the parameters. For example, time constants should be always positive values. Besides, the parameter bounds should be further tightened given any additional information and expert knowledge.

### B. Nonlinear Least Squares for Dynamic Systems

To solve the NLSQ problems (9-12), the stationary points must satisfy the first-order optimality condition:

$$\frac{\partial h(\boldsymbol{\theta})}{\partial \boldsymbol{\theta}} = -\sum_{k=1}^{N} \left(\frac{\partial \boldsymbol{y}(t_k)}{\partial \boldsymbol{\theta}}\right)^T (\boldsymbol{z}(t_k) - \boldsymbol{y}(t_k)) = \boldsymbol{0} \quad (13)$$

Starting from an initial estimate $\boldsymbol{\theta_0}$, a Newton iteration is used to solve the above Equation (13) and is expressed as follows:

$$\boldsymbol{\theta_{i+1}} = \boldsymbol{\theta_i} - \left[\left(\frac{\partial^2 h(\boldsymbol{\theta})}{\partial \boldsymbol{\theta}^2}\right)_i\right]^{-1} \left(\frac{\partial h(\boldsymbol{\theta})}{\partial \boldsymbol{\theta}}\right)_i \quad (14)$$

where $i$ is the iteration index.

The Hessian matrix of the objective function is given as follows:

$$\frac{\partial^2 h(\boldsymbol{\theta})}{\partial \boldsymbol{\theta}^2} = \sum_{k=1}^{N} \left(\frac{\partial \boldsymbol{y}(t_k)}{\partial \boldsymbol{\theta}}\right)^T \left(\frac{\partial \boldsymbol{y}(t_k)}{\partial \boldsymbol{\theta}}\right) \\ - \sum_{k=1}^{N} \left(\frac{\partial^2 \boldsymbol{y}(t_k)}{\partial \boldsymbol{\theta}^2}\right)^T (\boldsymbol{z}(t_k) - \boldsymbol{y}(t_k)) \quad (15)$$

Based on Equation (13-15), it can be seen that three types of computation are required for each iteration of the parameter update: 1) the system output response $\boldsymbol{y}(t_k)$ evaluated at $\boldsymbol{\theta_i}$; 2) the gradient of system output response $\frac{\partial \boldsymbol{y}(t_k)}{\partial \boldsymbol{\theta}}$ evaluated at $\boldsymbol{\theta_i}$; and 3) the Hessian matrix of system output response $\frac{\partial^2 \boldsymbol{y}(t_k)}{\partial \boldsymbol{\theta}^2}$ evaluated at $\boldsymbol{\theta_i}$.

The computation of model response $\boldsymbol{y}(t_k)$ only relies on the estimated parameters $\boldsymbol{\theta_i}$ and the state variable vector $\boldsymbol{x}(t_k)$. The latter one is obtained by simply integrating the state equation (10).

The computation of the gradient of system output response $\frac{\partial \boldsymbol{y}(t_k)}{\partial \boldsymbol{\theta}}$ requires more involved computation, especially for nonlinear systems. The analytical approach to obtain partial derivatives of the system equations (10-11) is not used in this paper for two reasons [23]. First, the analytical approach requires rederivation of the partial derivatives and additional implementation efforts if alternative model structures are used. Besides, for a system with discontinuous nonlinearities, it is difficult to define the derivatives analytically. The approach used in this paper is to approximate the derivative of system output response using the forward difference approximation [23]. This approach is a general method that can be applied to any nonlinear system, thus avoiding rederivation of the partial derivatives. The approximation is formulated as:

$$\left[\frac{\partial \boldsymbol{y}(t_k)}{\partial \boldsymbol{\theta}}\right]_{ij} \approx \frac{\tilde{y}_i(t_k) - y_i(t_k)}{\delta \theta_j}$$
$$\approx \frac{g_i(\tilde{\boldsymbol{x}}(t_k), \boldsymbol{u}(t_k), \tilde{\boldsymbol{\theta}}) - g_i(\boldsymbol{x}(t_k), \boldsymbol{u}(t_k), \boldsymbol{\theta})}{\delta \theta_j} \quad (16)$$
$$i = 1, \ldots, m; j = 1, \ldots, n$$

where $\delta \theta_j$ is a small perturbation of the $j$th entry of $\boldsymbol{\theta}$; $\tilde{\boldsymbol{\theta}}$ is the perturbed parameter vector corresponding to $\delta \theta_j$ and $\tilde{\boldsymbol{\theta}} = \boldsymbol{\theta} + \delta \theta_j \boldsymbol{e^j}$ (where $\boldsymbol{e^j}$ is a column vector having 1 in the $j$th row and 0 elsewhere); $\tilde{\boldsymbol{x}}(t_k)$ is the state variable vector after the perturbation; $\tilde{y}_i(t_k)$ is the $i$th output response after the perturbation. How $\tilde{y}_i(t_k)$ is computed is similar to $y_i(t_k)$, as explained previously, except that the parameter vector $\boldsymbol{\theta}$ is replaced by the perturbed parameter vector $\tilde{\boldsymbol{\theta}}$.

The Hessian matrix of the system output response is the most computationally expensive term to compute. One common approach is to neglect the second term in (15) based on the assumption that the residual $\boldsymbol{z}(t_k) - \boldsymbol{y}(t_k)$ is close to zero in a small-sized neighborhood of the solution $\boldsymbol{\theta}^*$. Therefore, substituting (13) and (15) into (14) yields the parameter update estimate as follows:

$$\Delta \boldsymbol{\theta_i} \approx \left(\sum_{k=1}^{N} \left(\frac{\partial \boldsymbol{y}(t_k)}{\partial \boldsymbol{\theta}}\right)_i^T \left(\frac{\partial \boldsymbol{y}(t_k)}{\partial \boldsymbol{\theta}}\right)_i\right)^{-1} \left(\sum_{k=1}^{N} \left(\frac{\partial \boldsymbol{y}(t_k)}{\partial \boldsymbol{\theta}}\right)_i^T \left(\boldsymbol{z}(t_k) - (\boldsymbol{y}(t_k))_i\right)\right) \quad (17)$$

IV. HYBRID BOX-CONSTRAINED LEVENBERG-MARQUARDT (H-BCLM) ALGORITHM

A. Levenberg-Marquardt Algorithm

The diesel generator model parameterization problem can be a rank-deficient NLSQ problem. $\frac{\partial \boldsymbol{y}(t_k)}{\partial \boldsymbol{\theta}}$ is a $m \times n$ matrix with $m < n$ because of the fact that the number of measurement points, such as frequency and voltage, are less than the number of the parameters to be estimated. Therefore, an inequality is derived as follows:

$$Rank\left[\left(\frac{\partial \boldsymbol{y}(t_k)}{\partial \boldsymbol{\theta}}\right)_i^T \left(\frac{\partial \boldsymbol{y}(t_k)}{\partial \boldsymbol{\theta}}\right)_i\right]$$
$$\leq min\left(Rank\left(\left(\frac{\partial \boldsymbol{y}(t_k)}{\partial \boldsymbol{\theta}}\right)_i^T\right), Rank\left(\left(\frac{\partial \boldsymbol{y}(t_k)}{\partial \boldsymbol{\theta}}\right)_i\right)\right)$$
$$\leq m < n$$

One can see that the $n \times n$ matrix $\left(\frac{\partial \boldsymbol{y}(t_k)}{\partial \boldsymbol{\theta}}\right)_i^T \left(\frac{\partial \boldsymbol{y}(t_k)}{\partial \boldsymbol{\theta}}\right)_i$ is a rank-deficient matrix. Because the input disturbance $\boldsymbol{u}(t)$ represents step load change and it lacks input signal richness [24], $\sum_{k=1}^{N} \left(\frac{\partial \boldsymbol{y}(t_k)}{\partial \boldsymbol{\theta}}\right)_i^T \left(\frac{\partial \boldsymbol{y}(t_k)}{\partial \boldsymbol{\theta}}\right)_i$ can also be a rank-deficient matrix, which is singular and noninvertible. In this case, the problem is a rank-deficient NLSQ problem. If conventional algorithms such as Gauss-Newton algorithm are applied to the problem [15], divergence will occur because of the singularity of $\sum_{k=1}^{N} \left(\frac{\partial \boldsymbol{y}(t_k)}{\partial \boldsymbol{\theta}}\right)_i^T \left(\frac{\partial \boldsymbol{y}(t_k)}{\partial \boldsymbol{\theta}}\right)_i$.

The LM algorithm, however, is robust when addressing rank-deficient problems [25] by introducing a trust region for $\Delta \boldsymbol{\theta}$. The objective function of the NLSQ problem (9) becomes

$$\min_{\Delta \boldsymbol{\theta}} \sum_{k=1}^{N} \left\|-\left(\frac{\partial \boldsymbol{y}(t_k)}{\partial \boldsymbol{\theta}}\right)_i^T \Delta \boldsymbol{\theta} + \left(\boldsymbol{z}(t_k) - (\boldsymbol{y}(t_k))_i\right)\right\|_2^2 \quad (18)$$
$$+ \lambda_i \|\Delta \boldsymbol{\theta}\|_2^2$$

where $\lambda_i$ is the LM parameter that controls both update search direction and step size at the $i$th iteration.

The normal equation of (18), which becomes the parameter update formula, is written as follows:

$$\Delta \boldsymbol{\theta_i} = \left(\sum_{k=1}^{N} \left(\frac{\partial \boldsymbol{y}(t_k)}{\partial \boldsymbol{\theta}}\right)_i^T \left(\frac{\partial \boldsymbol{y}(t_k)}{\partial \boldsymbol{\theta}}\right)_i + \lambda_i \boldsymbol{I}\right)^{-1} \left(\sum_{k=1}^{N} \left(\frac{\partial \boldsymbol{y}(t_k)}{\partial \boldsymbol{\theta}}\right)_i^T \left(\boldsymbol{z}(t_k) - (\boldsymbol{y}(t_k))_i\right)\right) \quad (19)$$

By adding the term $\lambda_i \boldsymbol{I}$, the LM algorithm is robust to the singularity of $\sum_{k=1}^{N} \left(\frac{\partial \boldsymbol{y}(t_k)}{\partial \boldsymbol{\theta}}\right)_i^T \left(\frac{\partial \boldsymbol{y}(t_k)}{\partial \boldsymbol{\theta}}\right)_i$. Note that a Gauss-Newton step is performed for $\Delta \boldsymbol{\theta_i}$ when $\lambda_i = 0$ while a gradient-descent step is performed under a large value of $\lambda_i$. For the implementation of choosing $\lambda_i$, the objective costs among the original $\lambda_i$, $a\lambda_i$ and $\frac{1}{a}\lambda_i (a < 1)$ are compared. What leads to the best cost minimization is chosen to be the final $\lambda_i$.

In this paper, the initial value for $\lambda$ is chosen to be 0.001 and $a$ is chosen to be 0.1. The choice of the initial value for $\lambda$ depends on the order of the magnitude of the matrix $\sum_{k=1}^{N} \left(\frac{\partial \boldsymbol{y}(t_k)}{\partial \boldsymbol{\theta}}\right)_i^T \left(\frac{\partial \boldsymbol{y}(t_k)}{\partial \boldsymbol{\theta}}\right)_i$. The choice of the initial value for $a$ to be 0.1 is somewhat arbitrary but proves effective in many NLSQ test cases considering dynamic systems [23].

B. Box-Constrained Levenberg-Marquardt Algorithm

Although the LM algorithm proves effective in solving rank-deficient NLSQ problems, it is only applicable to unconstrained problems. Therefore, a variant of the LM algorithm, BCLM, is proposed in this subsection to enable the LM algorithm to handle parameter constraints while maintaining the robustness of the algorithm. The advantages of the BCLM algorithm over the LM algorithm is twofold. *First*, the BCLM algorithm avoids any estimates which violate the physical limits of the parameters. For example, time constants could be set all greater than zero such that the estimates with non-positive time constants are eliminated. *Second*, from the perspective of practical use, the flexibility of the algorithm is improved by allowing experienced professionals who have prior knowledge of certain parameter ranges to tighten the constraints and narrow down the solution space.

Box-constrained transformation is the key technique that distinguishes the BCLM algorithm from the LM algorithm. The idea behind the BCLM algorithm is to transform the bounded variables into unbounded variables, and then the LM parameter update formula is applied to the unbounded variables. The unbounded variables are mapped into the original limits and transformed back to the bounded variables. To generalize the constraints defined in (12), three types of box constraints are redefined elementwise as follows:

*Type I:* $\underline{\theta_i} \leq \theta_i \leq \overline{\theta_i}$. A transformation from unbounded variables to bounded variables is shown as follows:

$$\theta_i = F_i(\beta_i) = \frac{\overline{\theta_i} - \underline{\theta_i}}{2} sin\frac{\pi}{2}\beta_i + \frac{\overline{\theta_i} + \underline{\theta_i}}{2} \quad (20)$$

where $\beta_i \in \mathbb{R}$. Since $sin\frac{\pi}{2}\beta_i \in [-1,1]$ for any $\beta_i$, $\theta_i$ will be always bounded by $[\underline{\theta_i}, \overline{\theta_i}]$. Its inverse transformation is written as follows:

$$\beta_i = F_i^{-1}(\theta_i) = \frac{2}{\pi} sin^{-1}\frac{2\theta_i - (\overline{\theta_i} + \underline{\theta_i})}{\overline{\theta_i} - \underline{\theta_i}} \quad (21)$$

Although the inverse transformation leads to an initial $\beta_i$ bounded by $[-1,1]$, it can go freely in $\mathbb{R}$ in the parameter update process.

*Type II:* $\underline{\theta_i} \leq \theta_i$. The transformation and its inverse transformation are written as follows:

$$\theta_i = F_i(\beta_i) = \underline{\theta_i} - 1 + \sqrt{\beta_i^2 + 1} \quad (22)$$

$$\beta_i = F_i^{-1}(\theta_i) = \sqrt{(\theta_i + 1 - \underline{\theta_i})^2 - 1} \quad (23)$$

*Type III:* $\theta_i \leq \overline{\theta_i}$. The transformation and its inverse transformation are written as follows:

$$\theta_i = F_i(\beta_i) = \overline{\theta_i} + 1 - \sqrt{\beta_i^2 + 1} \quad (24)$$

$$\beta_i = F_i^{-1}(\theta_i) = \sqrt{(\overline{\theta_i} + 1 - \theta_i)^2 - 1} \quad (25)$$

By substituting $\boldsymbol{\theta} = \boldsymbol{F}(\boldsymbol{\beta})$, the mapping function from $\boldsymbol{\beta}$ to $\boldsymbol{\Theta}$, into the NLSQ problem (9-12), one gets the following unconstrained NLSQ problem:

$$\min_{\boldsymbol{\beta}} h(\boldsymbol{\beta}) = \min_{\boldsymbol{\beta}} \frac{1}{2} \sum_{k=1}^{N} (\boldsymbol{z}(t_k) - \boldsymbol{y}(t_k))^T (\boldsymbol{z}(t_k) - \boldsymbol{y}(t_k)) \quad (26)$$

$$\dot{\boldsymbol{x}}(t_k) = \boldsymbol{f}[\boldsymbol{x}(t_k), \boldsymbol{u}(t_k), \boldsymbol{F}(\boldsymbol{\beta})], \boldsymbol{x}(t_0) = \boldsymbol{x_0} \quad (27)$$

$$\boldsymbol{y}(t_k) = \boldsymbol{g}[\boldsymbol{x}(t_k), \boldsymbol{u}(t_k), \boldsymbol{F}(\boldsymbol{\beta})] \quad (28)$$

Without changing the framework of the LM algorithm, the parameter update formula now updates the unconstrained parameter vector $\boldsymbol{\beta}$ instead of the bounded parameter vector $\boldsymbol{\theta}$ using

$$\Delta\boldsymbol{\beta_i} = \left(\sum_{k=1}^{N}\left(\frac{\partial \boldsymbol{y}(t_k)}{\partial \boldsymbol{\beta}}\right)_i^T \left(\frac{\partial \boldsymbol{y}(t_k)}{\partial \boldsymbol{\beta}}\right)_i + \lambda_i \boldsymbol{I}\right)^{-1} \\ \left(\sum_{k=1}^{N}\left(\frac{\partial \boldsymbol{y}(t_k)}{\partial \boldsymbol{\beta}}\right)_i^T \left(\boldsymbol{z}(t_k) - (\boldsymbol{y}(t_k))_i\right)\right) \quad (29)$$

According to the chain rule, the gradient of system output response in terms of $\boldsymbol{\beta}$ is written as follows:

$$\left(\frac{\partial \boldsymbol{y}(t_k)}{\partial \boldsymbol{\beta}}\right)_i = \left(\frac{\partial \boldsymbol{y}(t_k)}{\partial \boldsymbol{\theta}}\right)_i \left(\frac{\partial \boldsymbol{F}(\boldsymbol{\beta})}{\partial \boldsymbol{\beta}}\right)_i \quad (30)$$

where $\left(\frac{\partial \boldsymbol{y}(t_k)}{\partial \boldsymbol{\theta}}\right)_i$ is evaluated using (16) and $\left(\frac{\partial \boldsymbol{F}(\boldsymbol{\beta})}{\partial \boldsymbol{\beta}}\right)_i$ is evaluated using the analytical derivatives of $\boldsymbol{F}(\boldsymbol{\beta})$.

### C. Heuristic Algorithm for Initial Solution Search

As introduced in Section III, the convergence of the local search method is subject to the initial estimate, which is difficult to obtain from manufacturer datasheet for small-scale diesel generator. Therefore, a heuristic search algorithm Generalized Opposition-based Learning Genetic Algorithm (GOL-GA) is implemented in this paper to conduct global search within parameter bounds and produce an initial estimate for the local search method BCLM.

The fitness function for the GOL-GA is defined as follows:

$$\Gamma(\boldsymbol{\theta_k}) = 1/h(\boldsymbol{\theta_k}) \quad (31)$$

where $\boldsymbol{\theta_k}$ is the *k*th set of $\boldsymbol{\theta}$, and is also the *k*th chromosome in the population. Based on the fitness level, the Selection operation determines which chromosomes are the parent solutions for the Cross operation. The Crossover operation, which is used to generate new $\boldsymbol{\theta}$ using the parent solutions, is expressed as follows:

$$\begin{aligned}\theta_{k,i}^{new} &= (1-\lambda)\theta_{k,i} + \lambda\theta_{l,i} \\ \theta_{l,i}^{new} &= \lambda\theta_{k,i} + (1-\lambda)\theta_{l,i}\end{aligned} \quad (32)$$

where $\theta_{k,i}$ and $\theta_{l,i}$ are the *i*th parameter of the two parent solutions, $\boldsymbol{\theta_k}$ and $\boldsymbol{\theta_l}$, respectively; $\theta_{k,i}^{new}$ and $\theta_{l,i}^{new}$ are the *i*th parameter of the new solutions, respectively; $\lambda$ is a random real number within [0, 1]. The Mutation operation, which transforms the low-fitness solutions into new solutions using the GOL scheme [20]:

$$\theta_{m,i}^{new} = \begin{cases} \lambda(l_i^d + u_i^d) - \theta_{m,i}, if\ \theta_{m,i}^{new} \in [l_i^d, u_i^d] \\ rand(l_i^d, u_i^d), if\ \theta_{m,i}^{new} \notin [l_i^d, u_i^d] \end{cases} \quad (33)$$

$$l_i^d = \min_{m \in P}(\theta_{m,i}), u_i^d = \max_{m \in P}(\theta_{m,i}) \quad (34)$$

where $l_i^d$ and $u_i^d$ are the minimum and maximum of the *i*th parameter in the current population $P$.

The best solution generated by GOL-GA is used as the initial solution $\boldsymbol{\theta_0}$ for the BCLM algorithm. The BCLM algorithm is then applied to conduct local search using the parameter update formula (29).

### D. Overview of Model Parameterization Method

Fig. 4 shows the flowchart of the proposed model parameterization method using the H-BCLM algorithm, where the steps in **bold** mean that they involve running diesel generator dynamic model. The diesel generator dynamic model consists of two types of mathematical equations: state equations that include differential equations (1), (3), and (5), and observation equations that relate state variables to output response. The input profile for running the dynamic model is the load step profile, which is consistent with the load-step-change tests conducted in the field. A fourth-order Runge-Kutta method is used as the numerical simulation technique. The stopping criteria for both GOL-GA and BCLM consist of: (a) the iteration number reaches the maximum limit and (b) the relative cost difference reaches the threshold, such as $|(h_i - h_{i-1})/h_i| \leq 10^{-6}$.

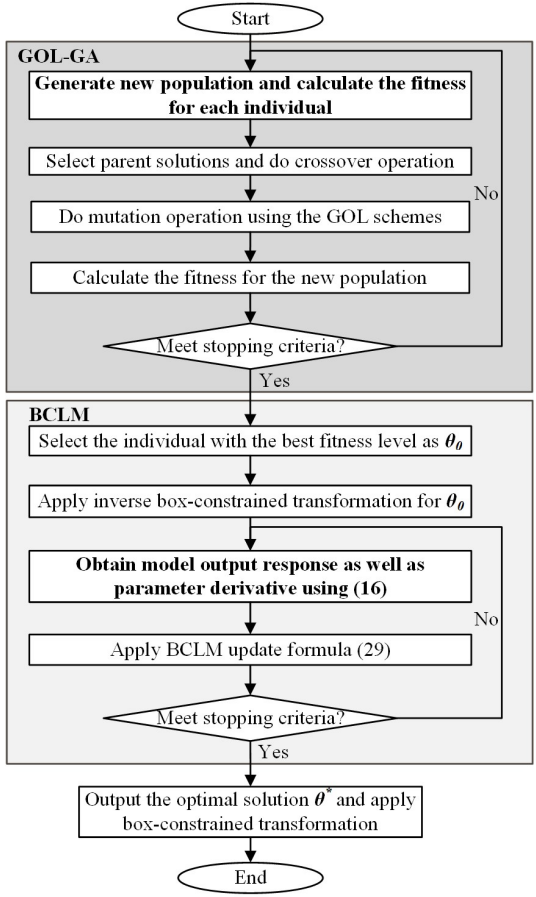

Fig. 4. Schematic of the proposed model parameterization method

As shown in Fig. 4, it is noted that the transformation and its inverse transformation are only applied to the model variables after and before the BCLM iteration. Its inverse transformation maps all the bounded parameters to an unbounded solution space before updating the parameter using (29). The transformation is then applied to map the solution from the unconstrained space back to the original space. By doing so, the BCLM method is able to handle parameters with bounds without changing the original LM algorithm structure. The obtained solutions are guaranteed to fulfil the original constrained NLSQ problem [26].

It should be remarked that the proposed hybrid model parameterization method is also applicable to diesel generators with other model structures such as converter-fed diesel engine generators [27, 28], as well as other microgrid component models such as solar and energy storage systems [28]. However, practical aspects should be paid attention to when this method is applied to other applications. For example, it is worth considering whether the resolution of the measurements needs to be changed or whether additional tests and measurement points are needed to complement the model parameterization process.

## V. CASE STUDIES

This section presents two case studies to demonstrate the effectiveness of the proposed model parameterization method. First, the benchmark test for H-BCLM is conducted against a diesel generator model with known parameters. Then, field testing measurements are used to validate the H-BCLM algorithm.

### A. Benchmark Case

The benchmark case tests the effectiveness of the H-BCLM algorithm, along with both the LM and the BCLM algorithms, against a diesel generator model with known parameters. An 50kVA, 400V diesel generator model is developed in MATLAB/Simulink®. The actual parameter values listed in

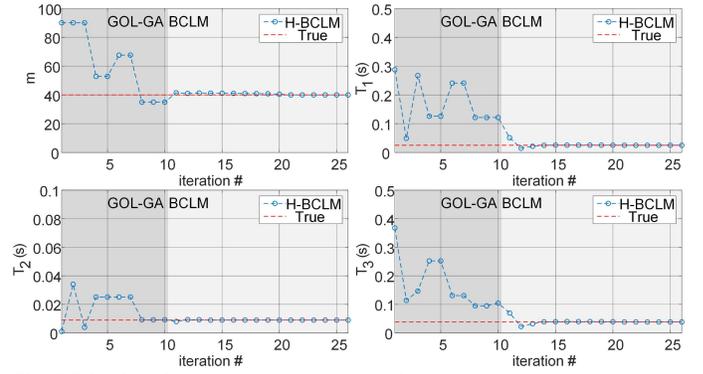

Fig. 5. Diesel engine parameter convergence in Case 4

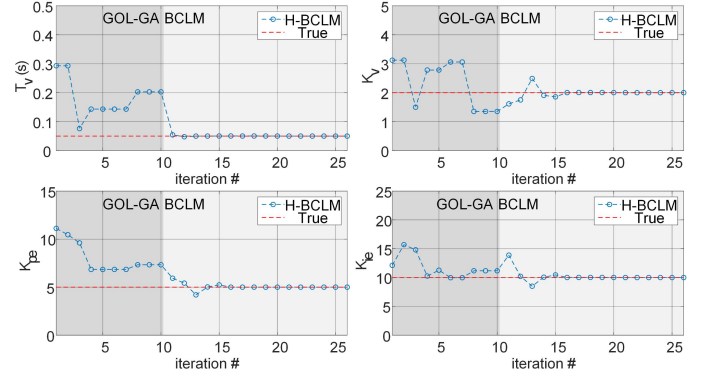

Fig. 6. Excitation system parameter convergence in Case 4

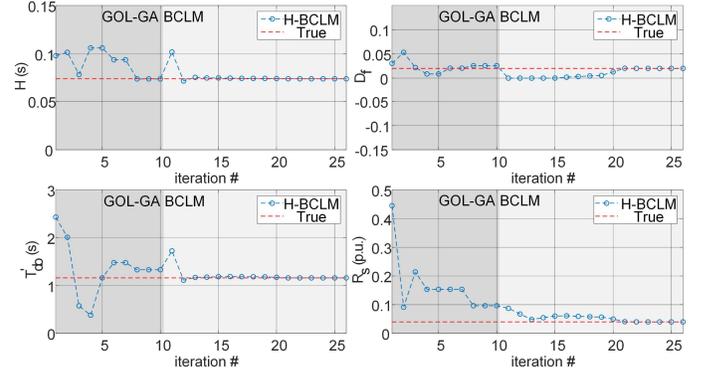

Fig. 7. Synchronous generator parameter convergence in Case 4

TABLE II
DIESEL GENERATOR BENCHMARK MODEL PARAMETERS AND TEST RESULTS

| | Symbol | $m$ | $T_1$ | $T_2$ | $T_3$ | $T_V$ | $K_V$ | $K_{pe}$ | $K_{ie}$ | $H$ | $D_f$ | $T'_{do}$ | $R_s$ |
|---|---|---|---|---|---|---|---|---|---|---|---|---|---|
| | Quantity | 40 | 0.025 | 0.009 | 0.038 | 0.05 | 2 | 5 | 10 | 0.074 | 0.020 | 1.16 | 0.04 |
| | Lower Bounds | 0 | 0 | 0 | 0 | 0 | 0 | 0 | 0 | 0.05 | 0 | 0 | 0 |
| | Upper Bounds | Inf | 0.5 | 0.5 | 0.5 | 0.5 | Inf | Inf | Inf | 0.15 | Inf | Inf | Inf |
| Case 1 | Init | 120 | 0.125 | 0.045 | 0.19 | 0.25 | 10 | 25 | 50 | 0.14 | 0.1 | 5 | 0.2 |
| | LM | 40 | 0.025 | 0.009 | 0.038 | 0.05 | 2 | 5 | 10 | 0.074 | 0.020 | 1.16 | 0.04 |
| | BCLM | 40 | 0.025 | 0.009 | 0.038 | 0.05 | 2 | 5 | 10 | 0.074 | 0.020 | 1.16 | 0.04 |
| Case 2 | Init | 80 | 0.25 | 0.09 | 0.4 | 0.5 | 20 | 50 | 100 | 0.14 | 0.2 | 2.3 | 0.4 |
| | LM | 104 | -5672 | 0.012 | -14689 | 0.05 | 3.5 | 2.9 | 5.7 | 0.088 | -0.59 | 1.16 | 0.04 |
| | BCLM | 40 | 0.025 | 0.009 | 0.038 | 0.05 | 4.1 | 2.4 | 4.8 | 0.074 | 0.020 | 1.16 | 0.04 |
| Case 3 | Init | 400 | 0.25 | 0.09 | 0.4 | 0.5 | 20 | 50 | 100 | 0.14 | 0.2 | 5 | 0.4 |
| | LM | 0.44 | 0.002 | 20.01 | 20.01 | 0.76 | 13 | 17 | 80.8 | 0 | 0 | 0.56 | 0.45 |
| | BCLM | 300 | 0.006 | 0.399 | 0.490 | 0.05 | 2.8 | 3.6 | 7.2 | 0.071 | 4765 | 1.16 | 0.04 |
| Case 4 | H-BCLM | 40 | 0.025 | 0.009 | 0.038 | 0.05 | 2 | 5 | 10 | 0.074 | 0.020 | 1.16 | 0.04 |

*Machine reactance initial estimates are: $X_d$ = 3.79, $X_q$ = 2.12, $X'_d$ = 0.342.

Table II are selected based on a combination of online sources [27] and Simulink® preset model for synchronous machines. The fixed parameters contain all the reference values, which are at 1.0 p.u. All the machine reactances have the initial estimates from manufacturer datasheet with tight bounds, shown in Table II. Two load-step-change tests are conducted: a load step-up test from 30% to 80% of the rated power, and a load step-down test from 80% to 30%. The frequency and voltage measurements at the generator terminal are the input to the H-BCLM algorithm.

Four cases are presented in Table II. From Case 1 to 3, different initial estimates are considered for both LM and BCLM algorithms while in Case 4, the H-BCLM algorithm is applied without initial estimates. Case 1 considers "reasonable" initial estimates, which are set five times as the true values. Case 2 and Case 3 consider "unreasonable" initial estimates, which are selected to be no less than twice as the true values, and more than half of which are more than ten times of the true values. The prior knowledge is converted into parameter bounds as box constraints that are listed in Table II considering: 1) all the control gains are larger than 0; 2) the friction factor and the stator resistance are larger than 0; 3) all the time constants are larger than 0; and 4) the constraints for the inertia constant are set based on the generator rated capacity.

The final estimates are also shown in Table II. In Case 1, all the parameters converge to the exact actual values using both LM algorithm and BCLM algorithm. With less accurate initial estimates than Case 1, it can be observed in Case 2 that the LM algorithm performs poorly and fails to accurately estimate the parameters. Diesel engine time constants $T_1$ and $T_3$, and friction factor $D_f$ reach infeasible ranges. Unlike the LM method, the estimated parameters converge to the actual values using the BCLM algorithm in Case 2 after the constraints are imposed. In Case 3, diesel engine time constants $T_1$ and $T_2$, friction factor $D_f$ and inertia constant $H$ end with impractical estimates using the LM algorithm. Meanwhile, the BCLM algorithm fails to obtain the global optimal solution, indicating the initial solution is far away from the optimum.

The parameter estimation convergence for Case 4 is shown in Fig. 5, 6 and 7. Without prior estimates, all the parameters converge to the actual values using the H-BCLM algorithm. The heuristic algorithm GOL-GA and the BCLM algorithm are color coded to show both global and local search process.

In terms of computation complexity, the H-BCLM algorithm always converges in less than 60 iterations in total, including 10 GOL-GA iterations. The meantime to solve using the H-BCLM algorithm is 30000 seconds given a 2.50 GHz Core-i7 16GB RAM machine. The main computation lies in running numerical simulation using the 4th Order Runge-Kutta method. Depending on test duration and time step, the computation time varies in a wide range. In case where test duration is selected to be 30 seconds with a time step of 100 microseconds, the computation time has not been regarded as a major issue.

Overall, the BCLM algorithm shows good performance and robustness in accurately estimating the parameters over the LM algorithm. By incorporating the prior knowledge into the BCLM algorithm, the solution space is reduced, and any infeasible estimate is eliminated. However, both algorithms are subject to initial solution selection. The H-BCLM algorithm is successful in obtaining the optimum regardless of initial solutions.

### B. Model Parameterization for a Small Diesel Generator

The second case provides validation results for matching field measurement from an actual 16-kVA diesel generator running in a commercial microgrid site [28]. The load-step-change tests for the diesel generator are done using an electric load emulator. A high-resolution digital recorder is installed at the terminal of the diesel generator to record the three-phase output voltage and current, with the sampling frequency of 1kHz. The raw data is post-processed using low-pass filters, which are used for filtering out the high-frequency measurement noise, and phase-locked loops, which are applied to extract the dynamics of frequency and voltage magnitude.

The effectiveness of the proposed H-BCLM is compared to three algorithms, which were selected from literature reviews: 1) the constrained Gauss-Newton with linear search (CGN-LS) [15, 23]; 2) the state-of-the-art TRR [17]; and 3) the state-of-the-art DPSO-LS [19]. Performance comparison is provided under the same parameter bounds and stopping criteria. Like the benchmark case in Section V-A, the machine reactance is tightly bounded with the initial estimates from manufacturer datasheet.

The extracted solutions from field data are provided in Table III as well as the accuracy of matching errors between field measurements and simulated results. The CGN-LS experiences divergence issues. The reason is that, with the limited measurement points, the diesel generator model parameterization is a rank-deficient NLSQ problem, leading to an invalid Gauss-Newton step. The eigenvalues of the left matrix for the Gauss-Newton step (17) is computed. A small eigenvalue in the order of $10^{-12}$ is found as follows, which validates the rank-deficiency of the original Hessian approximation:

$$eig\left[\sum_{k=1}^{N}\left(\frac{\partial \boldsymbol{y}(t_k)}{\partial \boldsymbol{\theta}}\right)_i^T \left(\frac{\partial \boldsymbol{y}(t_k)}{\partial \boldsymbol{\theta}}\right)_i\right]$$
$$= [1.15e^8 \quad 1.81e^7 \quad 4.75e^6 \quad 3.23e^6 \quad 1.06e^6 \quad 2.23e^5 \quad 1.55e^5$$
$$\quad 4.00e^3 \quad 284.81 \quad 2.52 \quad -1.19e^{-12}]$$

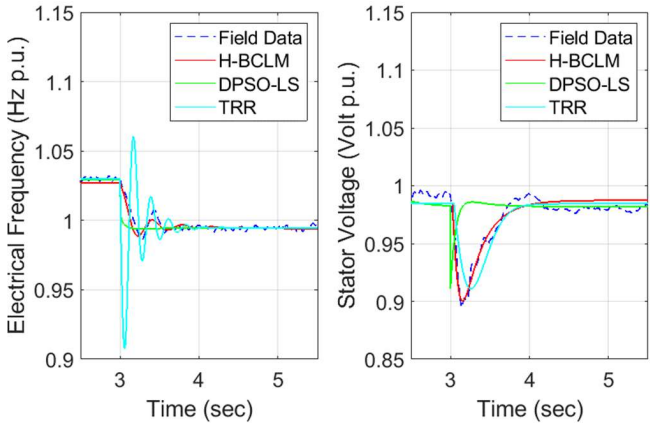

Fig. 8. Optimal fit for load step-up test

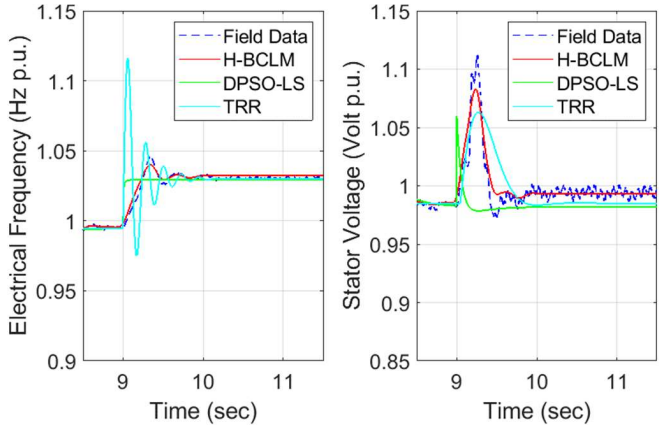

Fig. 9. Optimal fit for load step-down test

Fig. 8 and Fig. 9 show the fitted frequency and voltage responses using H-BCLM and another two algorithms. While all the algorithms capture the steady-state responses correctly, they result in different levels of agreement with the dynamics observed in the field measurements. The TRR shows decent performance in capturing voltage dynamics while performing poorly in matching frequency dynamics. The cause for the poor performance is similar to BCLM in Section V-A, that is, the TRR algorithm itself is a local search algorithm whose performance is highly dependent on the selection of initial solution. However, the DPSO-LS performs the opposite way, leading to a small deviation in matching frequency dynamics and a large deviation of matching voltage dynamics. The reason why DPSO-LS exhibits degraded performance is because it is still a pure heuristic algorithm that ignores gradient information for dynamic system even though a similar GOL is utilized.

TABLE III
EXTRACTED PARAMETERS OF DIESEL GENERATOR MODEL

| Estimated Parameters | CGN-LS | TRR | DPSO-LS | H-BCLM |
|---|---|---|---|---|
| $m$ | | 15.838 | 19.042 | 15.479 |
| $T_1$ | | 0.042 | 0.164 | 1.494 |
| $T_2$ | | 0.119 | 0.0008 | 0.001 |
| $T_3$ | | 0.038 | 0.166 | 1.486 |
| $T_V$ | | 0.076 | 0.0001 | 0.056 |
| $K_V$ | Solution | 0.882 | 20.201 | 0.491 |
| $K_{pe}$ | Divergence | 7.042 | 5.165 | 16.462 |
| $K_{ie}$ | | 22.437 | 16.470 | 56.847 |
| $H$ | | 0.086 | 0.057 | 0.150 |
| $D_f$ | | 0.019 | 0.186 | 0.230 |
| $T'_{do}$ | | 0.978 | 4.792 | 0.837 |
| $R_s$ | | 0.137 | 0.354 | 0.094 |
| RMSE_f | NAN | 0.0183 | 0.0060 | **0.0025** |
| RMSE_V | NAN | 0.0144 | 0.0268 | **0.0076** |

Among all, the proposed H-BCLM algorithm generates a high-quality solution that fits both frequency and voltage responses and outperforms other selected algorithms. The small matching errors between model response and field measurement is mainly due to the simplification of the model structure and the lack of information on the measurement filters.

## VI. CONCLUSION

In this paper, a robust H-BCLM algorithm is proposed to parameterize the nonlinear diesel generator model for microgrid dynamic simulation. The uniqueness of the proposed method is three-fold. First, the proposed approach enables one to simultaneously generate a complete set of diesel generator model parameters without the need of knowing any initial estimate. Second, the H-BCLM algorithm improves the quality of the solutions by allowing the integration of expert knowledge and eliminating the invalid solutions with no physical meanings. Last but not least, the H-BCLM algorithm offers a robust approach to address the diesel generator model parameterization problem, which is identified as a rank-deficient NLSQ problem considering only load-step-change tests with limited measurement points. Compared with three existing state-of-the-art algorithms, the proposed H-BCLM algorithm always converges and reduces the matching errors by 50%. Our future work will focus on applying the proposed method to other applications such as solar and battery inverters.

## ACKNOWLEDGMENT

The authors would like to thank Dr. Wente Zeng and Mr. Stanislas DE Crevoisier D' Hurbache with Total S.A. for their support of this research project.

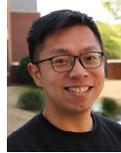

**Qian Long** (S'17) received the B.S. degree in electrical engineering from China Agricultural University, Beijing, China, in 2014, and both the M.S. degree and the Ph.D. degree in electrical engineering from North Carolina State University, Raleigh, NC, USA, in 2016 and 2020, respectively. His research interests are in the areas of modeling, operation, control and optimization of hybrid energy systems.

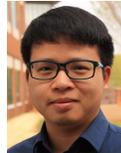

**Hui Yu** (S'17) received the Bachelor and Master degrees in electrical engineering from Huazhong University of Science and Technology, Wuhan, China, in 2013 and 2016, respectively. He is currently working toward the Ph.D. degree at the Future Renewable Electric Energy Delivery and Management (FREEDM) Systems Center, North Carolina State University, Raleigh, NC, USA. His research interests include power electronics converter modeling, design, and control in microgrids and energy storage systems. He was a recipient of the Outstanding Presentation Award of the 2018 IEEE Applied Power Electronics Conference.

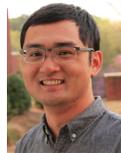

**Fuhong Xie** (S'16) received the B.S. and M.S. degree in electrical engineering from Harbin Institute of Technology, Harbin, China, in 2014 and 2016, respectively. Now he is pursuing the Ph.D. degree at the North Carolina State University, NC, USA. His research interests are in the theoretical and algorithmic studies in microgrid modeling, control, and real-time hardware-in-the-loop simulation.

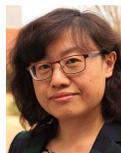

**Ning Lu** (M'98--SM'05) received the B.S. degree in electrical engineering from the Harbin Institute of Technology, Harbin, China, in 1993, and the M.S. and Ph.D. degrees in electric power engineering from Rensselaer Polytechnic Institute, Troy, NY, USA, in 1999 and 2002, respectively. Dr. Lu is currently a professor with the Electrical and Computer Engineering Department, North Carolina State University. She was with Pacific Northwest National Laboratory. Her current research interests include microgrid modeling and control, power distribution systems operation and planning, and power system data analysis.

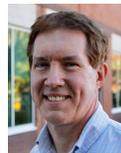

**David Lubkeman** (F'15) received the Ph.D. degree in electrical engineering from Purdue University with an emphasis in power systems engineering. He is also a Licensed Professional Engineer. Dr. Lubkeman is currently a Research Professor at North Carolina State University. His research interests include electric power distribution system analysis, distribution automation, distribution management systems, renewable energy integration and microgrid applications.